\documentstyle[12pt]{article}
\begin{document}
%%%%%%%%%%%%%%%%%%%%%%%%%%%%%%%%%%%%%%%%%%%%%%%%%
\def\D{{\cal D}}
\def\R{{{\rm I} \! {\rm R}}}
\newcommand{\dt}{\displaystyle}
\newcommand{\btup}{{\bigtriangleup}}
\newcommand{\id}{{\rm id}}
\newcommand{\N}{{\bf\rm N}}
\newcommand{\I}{{\bf\rm I}}
\newcommand{\ctg}{{\rm ctg}}
\newcommand{\e}{{\rm e}}
\newcommand{\mod}{{\rm mod}}
\newcommand{\df}{\stackrel{df}{=}}
%%%%%%%%%%%%%%%%%%%%%%%%%%%%%%%%%%%%%%%%%%%%%%%%%%
%%%%%%%%%%%%%%%%%%%%%%%%%%%%%%%%%%%%%%%%%%%%%%%%%%
\vspace*{2cm}
\begin{center}
 \Huge\bf
Solution of  the vacuum Einstein  equations
in Synthetic Differential Geometry of Kock-Lawvere

\vspace*{0.25in}

\large

Alexander K.\ Guts, Artem A. Zvyagintsev
\vspace*{0.15in}

\normalsize

Department of Mathematics, Omsk State University \\
644077 Omsk-77 RUSSIA
\\
\vspace*{0.5cm}
E-mail: guts@univer.omsk.su  \\
\vspace*{0.5cm}
Auguest 30, 1999\\
\vspace{.5in}
ABSTRACT
\end{center}

The topos theory is a theory which is used for deciding
a number of problems of theory of relativity, gravitation
and quantum physics.
It is known that topos-theoretic geometry can be successfully developed
within the framework of Synthetic Differential Geometry of
Kock-Lawvere (SDG), the models of which are serving the toposes,
i.e. categories possessing many characteristics of traditional Theory
of Sets. In the article by using ideas SDG, non-classical spherically
symmetric solution of
the vacuum Einstein equations is given.

\newpage

\setcounter{page}{1}

%%%%%%%%%%%%%%%%%%%%%%%%%%%%%%%%%%%%%%%%%%%%%%%%%%%%%%%%%%%%%%%%%%%%

Theoretical physics always tended  operatively to use new ideas coming
up from the mathematics. So it is not wonderful that new
topos-theoretic mathematics \cite{Goldblatt, Kock} was immediately
called to deciding a number of problems of theory of relativity
and gravitation \cite{Guc, Tri, Guc1, Grink, GG} and quantum
physics \cite{Ish}. Formally, for instance, it is
suitable to develop the topos-theoretic geometry within the
framework of Synthetic Differential Geometry of Kock-Lawvere \cite{Kock}
(further for brevity we write SDG), models of which are serving
toposes, i.e. categories possessing many characteristics of traditional
Theory of Sets. Last theory was the basis of mathematics of the XX century.
In the article by using ideas SDG, non-classical solution of
the vacuum Einstein  equations is given.

\section{Intuitionistic theory of gravitation}

Synthetic  Differential Geometry of Kock-Lawvere \cite{Kock}
is built on the base of change the field of real numbers ${\R}$ on
commutative ring ${\bf R}$, allowing to define on him differentiation,
integrating and "natural numbers". It is assumed that there exists
$D$ such that $D =\{ x\in {\bf R}~|~x^2=0 \}$ and that following
the Kock-Lawvere axiom is held:

\begin{quote}
for any $g:D\rightarrow {\bf R}$ it exist
the only $a, b\in{\bf R}$ such that for any $ d\in D ~~ g(d)=a+d\cdot b $.
\end{quote}

This means that any function in given geometry is differentiable, but
"the law of excluded middle" is false. In other words,
intuitionistic logic acts  in SDG. But on this way one is possible
building an intuitionistic theory of gravitation in analogy with the
General theory of Relativity of Einstein \cite{Guc1,Grink,GG}.
The  elements of $d\in D$ are called infinitesimals, i.e.
infinitesimal numbers.
On the ring ${\bf R}$ we can look as on the field of real numbers $\R$
complemented by infinitesimals.

The  vacuum Einstein  equations  in SDG in space-time ${\bf R}^4$ 
can be written with {\it nonzero}
tensor of the energy. For instance,
$$
R_{ik}-\frac{1}{2}g_{ik}(R-2\Lambda) =
\frac{8\pi G}{c^2} \ d u_iu_k, \eqno(1)
$$
where density of matter $d\in D$ is arbitrarily taken infinitesimal
\cite{Guts1}. For infinitesimals are holding relations which are impossible
from standpoints of classical logic: $d\neq 0$, $d\leq ~0$ $\& \ d\geq 0$
and $-\epsilon < d < \epsilon $ for any $\epsilon\in {\bf R}, \epsilon>0$.
Such non-classical density of vacuum matter will consistent with
zero in  right part of the Einstein's equations
in the case of the vacuum in classical General theory of Relativity.
For this one is sufficiently to consider SDG in so named
well-adapted models, in which we can act within the framework of
classical logic. For instance, in smooth topos
${\bf Set}^{\bf L_{op}}$,
where ${\bf L}$ category $C^\infty$-rings \cite{Moerdijk},
the equations (1) at stage of locus $\ell A=\ell(C^\infty (\R^n)/I)$,
$I$ is a certain ideal of  $C^{\infty}$-smooth functions 
from $\R^n$ to $\R$,
have the form
$$
R_{ik}(a)-\frac{1}{2}g_{ik}(a)(R(u)-2\Lambda(a))
= \frac{8\pi G}{c^2} \ d(a) u_i(a)u_k(a)\  mod\ I, \eqno(2)
$$
where $a\in \R^n$ in parenthesises shows that we have
functions, but at stage ${\bf 1}=\ell(C^\infty(\R)/\{a\})$,
equations (2) take a classical form with null
(on $ mod\ \{a\}$) tensor of the energy.

Note that an event $x $ of the space-time ${\bf R}^4$ at stage
        $\ell A$  is the class of
        $C^{\infty}$-smooth vector functions
        $(X^0(a),X^1(a),X^2(a),X^3(a)):\R^n\rightarrow  \R^4$,
where each function     $X^i(a)$ is taken by $mod \ I$.
     The argument $a\in\R^n$ is some "hidden" parameter corresponding
        to the stage $\ell A$. Hence it follows that at stage of real
        numbers ${\bf R}=\ell C^{\infty}(\R)$ of the topos 
under consideration  an event $x$ is described by just a
        $C^{\infty}$-smooth vector function
        $(X^0(a),X^1(a),X^2(a),X^3(a)), a\in \R$.
	At stage of ${\bf R}^2=\ell C^{\infty}(\R^2)$
        an event $x$ is 2-dimensional surface, i.e. a {\it string}.
      The classical four numbers $(x^0,\ x^1,\ x^2,\ x^3)$, the
coordinates of the event $x$,
        are obtained at the stage
${\bf 1}=\ell C^{\infty}(\R^0)= \ell C^{\infty}(\R)/\{a\}$
    (the ideal $\{a\}$ allows one to identify functions if their values at
        $0$ coincide), i.e., $x^i=X^i(0), i=0,1,2,3$.

\section{Spherically symmetrical field}

We have the Einstein equations describing the gravitational field
created by certain material system
$$
R_{ik}-\dt\frac{1}{2}g_{ik}(R-2\Lambda)=\kappa T_{ik}
$$
Here $R_{ik}=R_{ilk}^l,~R=g^{ik}R_{ik},~\kappa=8\pi G/c^4$.

Consider case, when gravitational field possesses a central symmetry.
Central symmetry of field means that interval
of space-time can be taken in the form
$$
ds^2=\e^{\nu (r,t)}dt^2-
e^{\lambda (r,t)}dr^2-r^2(d\theta^2 +\sin^2\theta\cdot d\varphi^2)
$$
Note that such type of metric does not define else
choice of time coordinate by unambiguous image:
given metric can be else subject to any transformation
of type $t=f(t^\prime)$ not containing $r.$

All calculations are conducted also as in the classical case.
Herewith we consider components of metric tensor by invertible values in
${\bf R}$. For the  Christoffel coefficients we have
the usual formula
$$
\Gamma_{kl}^i=\dt\frac{1}{2} g^{im}(\dt\frac{\partial g_{mk}}{\partial x^l}
+ \dt\frac{\partial g_{ml}}{\partial x^k} - \dt\frac{\partial g_{kl}}
{\partial x^m}).
$$
Hence we have
$$
\Gamma_{11}^1=\dt\frac{\lambda^{\prime}}{2},
\Gamma_{10}^0=\dt\frac{\nu^{\prime}}{2},
\Gamma_{33}^2=-\sin\theta \cos\theta,
\Gamma_{11}^0=\dt\frac{\dot{\lambda}}{2}\e^{\lambda - \nu},
\Gamma_{22}^1=-r\e^{-\lambda},
\Gamma_{00}^1=\dt\frac{\nu'}{2}\e^{\nu- \lambda},
$$
$$
\Gamma_{12}^2= \Gamma_{13}^3 =1/r ,\Gamma_{33}^3=\ctg\theta,
\Gamma_{00}^0=
\dt\frac{\dot{\nu}}{2},\Gamma_{10}^1=\dt\frac{\dot{\lambda}}{2},
\Gamma_{33}^1=-r \sin^2\theta\e^{-\lambda}.
$$

Here the prime means differentiation with respect to  $r,$
and dot means differentiation with respect to $t.$

Tensor of Ricci  is also calculated with help of  known formula
$$
R_{ik}=\frac{\partial\Gamma_{ki}^l}{\partial x_l}
-\frac{\partial\Gamma_{il}^l}{\partial x_k} +
\Gamma_{ik}^l\Gamma_{lm}^m-\Gamma_{il}^m\Gamma_{km}^l.
$$

The Einstein's equations have the form:
\setcounter{equation}{2}
\begin{equation}
{-\e^{-\lambda}(\dt\frac{\nu^{\prime}}{r} + \dt\frac{1}{r^2}) +
\dt\frac{1}{r^2} - \Lambda =\kappa T_1^1}
\end{equation}

\begin{equation}
{-\dt\frac{1}{2}\e^{-\lambda}(\nu^{\prime\prime}
+ \dt\frac{{\nu^{\prime}}^2}{2} + \dt\frac{\nu^{\prime}
- \lambda^{\prime}}{r} - \dt\frac{\nu^{\prime}\lambda^{\prime}}{2})
 + \dt\frac{1}{2}\e^{-\nu}(\ddot{\lambda} + \dt\frac{\dot{\lambda}^2}{2}
- \dt\frac{\dot{\nu}\dot{\lambda}}{2}) - \Lambda =\kappa T_2^2=\kappa T_3^3}
\end{equation}

\begin{equation}
{-\e^{-\lambda}(\dt\frac{1}{r^2} - \dt\frac{\lambda^{\prime}}{r}) +
\dt\frac{1}{r^2} - \Lambda =\kappa T_0^0}
\end{equation}

\begin{equation}
{-\e^{-\lambda}\dt\frac{\dot{\lambda}}{r} =\kappa T_0^1}
\end{equation}

Equation (4), as it is well known \cite{Syng}, is corollary of equations
(3), (5), (6) and the law of conservation
\begin{equation}
{T^{ik}_{\ \ ;k}=0.}
\end{equation}
So further the  equation (4) will be omited.

{\bf 2.1. Field of vacuum}. Consider now important example
of gravitational field in the vacuum. For this we will take
$T_k^i = c^2\rho u^iu_k,$ i.e. tensor of the energy of dust matter.
Here $\rho$ is density of dust in the space which will consider
further constant value.  Suppose  that dust is described in
coordinate system in which
$u_i=(\e^{-\frac{\nu}{2}},0,0,0),$
$u^k=g^{ik}u_i=(\e^{\frac{\nu}{2}},0,0,0).$
So $T_0^0=c^2\rho,~T_1^1=T_2^2=T_3^3=0$ and equations (3),(5),(6)
will take following forms

\begin{equation}{-\e^{-\lambda}(\dt\frac{\nu^{\prime}}{r}
+ \dt\frac{1}{r^2}) +
\dt\frac{1}{r^2} - \Lambda =0
}\end{equation}                                %6

\begin{equation}{-\e^{-\lambda}(\dt\frac{1}{r^2}
- \dt\frac{\lambda'}{r}) +
\dt\frac{1}{r^2} - \Lambda =\kappa c^2\rho           %8
}\end{equation}

\begin{equation}{-\e^{-\lambda}\dt\frac{\dot{\lambda}}{r} =0
}\end{equation}                                      %9

By using form of tensor $T_k^i$ and equation (10)
we   rewrite the equation (7) as follows
\begin{equation}{
\rho\nu^{\prime} =0
}
\end{equation}    %10

Try to solve equations (8)-(10) using equation (11).
From equation (10) it follows that $\lambda(r,t)=\lambda(r),$ i.e.
$\lambda$ does not depend on coordinate $t.$ As far as
$\rho$ and $\Lambda$ are constants, the equation (9) can be easy
integrated. Really, by taking  $\e^{-\lambda}$ for $u,$ we get
\begin{equation}{
u^{\prime} r + u=1 - (\Lambda + \kappa c^2\rho)r^2
}
\end{equation}                          %12
Solution of uniform equation $u^{\prime} r + u$  has form
$u=\dt\frac{A}{r},$ where $A=const.$ Thereby, for non-uniform
equation we will get
$u=\dt\frac{A(r)}{r}.$ Substituting this in (12), we have for $A(r)$:
$$
A^\prime (r)=1 - (\Lambda + \kappa c^2\rho)r^2
$$
Solution of this equation is function
$$
A(r)=r - \dt\frac{(\Lambda + \kappa c^2\rho)r^3}{3} + C.
$$
Thence
$$
u(r)= 1 - \dt\frac{(\Lambda + \kappa c^2\rho)r^2}{3} + \dt\frac{C}{r}
$$
or
\begin{equation}{
\e^{-\lambda}= 1 - \dt\frac{(\Lambda + \kappa c^2\rho )r^2}{3}
+ \dt\frac{C}{r}
}
\end{equation}       %13
Here $C$ is a constant of integrating.

Hereinafter, to find an expression for $\nu,$
we need integrate an equation (8).
But for the beginning we consider an equation (11).

Notice that $\rho=d,~ \nu=d$ under any $d\in D$ are its solutions.
Thereby, from existence of such objects
as $D,~D_2,~D(2),~\btup$ and etc \cite{Kock}, it follows that except
classical its solutions
$$
(\rho=0 \ \& \ \nu\not=0) \vee (\nu=0 \ \& \ \rho\not=0)
 \vee (\rho=0\ \& \ \nu=0)
$$
there exist and the others, non-classical one.
For example, $\rho$ and $\nu$ that are inseparable from
the zero ($x$ is separable from the zero, if there exists
a natural number $n$
such that $(1/n) < x \ \vee \ x < -(1/n)$) . The First from classical cases
above gives
well-known  the   classical Schwarzschild solution.

Consider non-classical case of deciding an equation (11),
when both values $\rho$ and
$\nu$ are simultaneously inseparable from the zero.
Substituting (13) in (8) and considering (11) we get
\begin{equation}{
\dt\frac{\nu^{\prime}}{r}(1+\dt\frac{C}{r} - \dt\frac{\Lambda r^2}{3})+
\dt\frac{2}{3}\Lambda - \dt\frac{\kappa c^2\rho}{3} + \dt\frac{C}{r^3}=0
}
\end{equation}   %14
Thence easy notice that $\dt\frac{2}{3}\Lambda + \dt\frac{C}{r^3} $
is inseparable from the zero. Besides, when considering this expressions
in topos ${\bf Set}^{\bf L_{op}}$ at stage {\bf 1} this expression
becomes an equal to zero that is possible in that case only, when
and $\Lambda$ and $C$ at this stage are zero.
Thereby, we conclude that $C$ and $\Lambda$ are also inconvertible,
but hence and $\dt\frac{C}{r} - \dt\frac{\Lambda r^2}{3}$
is inconvertible.
By using now (11), we will transform (14) to the form

$$
\nu^{\prime} (1+\dt\frac{C}{r} - \dt\frac{\Lambda r^2}{3}
+ \dt\frac{\kappa c^2\rho r^2}{6})= \dt\frac{\kappa c^2\rho r}{3} -
\dt\frac{2}{3}\Lambda\cdot r - \dt\frac{C}{r^2}
$$

or, that is equivalent,

\begin{equation}{
\nu^{\prime} =\dt\frac {\dt\frac{\kappa c^2\rho r}{3} -
\dt\frac{2}{3}\Lambda\cdot r - \dt\frac{C}{r^2}}{1+\dt\frac{C}{r}
 - \dt\frac{\Lambda r^2}{3}
+ \dt\frac{\kappa c^2\rho r^2}{6}},
}
\end{equation}   %15

Deciding equation (15) we find that
$$
\nu= \ln \left|1+ \dt\frac{C}{r} - \dt\frac{\Lambda r^2}{3}
+ \dt\frac{\kappa c^2\rho r^2}{6}\right| + f(t),
$$
where $f(t)$ is arbitrary function that is depending only on
coordinate $t.$ On the strength of that that we left for itself
else possibility of arbitrary transformation of time  $t=g(t^\prime),$
which is equivalent addition to $\nu$ an arbitrary functions of
time, $f(t)$ can always be made to be equal to zero.
Hence, not limiting generalities, it is possible to consider that

\begin{equation}{
\nu= \ln \left|1+ \dt\frac{C}{r} - \dt\frac{\Lambda r^2}{3}
+ \dt\frac{\kappa c^2\rho r^2}{6}\right|}
\end{equation}     %16

Substituting these values for $\lambda$ and $\nu$ in expression
for $ds^2,$ we get that
$$
ds^2= \left(1+ \dt\frac{(\kappa c^2\rho -2\Lambda ) r^2}{6} +
 \dt\frac{C}{r}\right)dt^2
-\frac{dr^2}
{1 - \dt\frac{(\Lambda + \kappa c^2\rho )r^2}{3} + \dt\frac{C}{r} }-
$$
$$
- r^2(d\theta^2 + \sin^2\theta\cdot d\varphi^2 ) \eqno(17)
$$
This metric can be called the  non-classical Schwarzschild  solution
of the Einstein equations

Suppose  that gravitational field has no singularity in all space.
This means that metric has no singularity in $r=0$. So we shall
consider that $C$ is  zero. Coming from this and multiplying the right
and left parts of equations (14) on $\rho,$ we get that
$2\Lambda\rho = \kappa c^2\rho^2$ and, besides, $\Lambda$ is inconvertible
value of ring ${\bf R}.$

% Thence follows that density of matter $\rho$ is inconvertible value.

In other words, matter has non-classical density,
and its gravitational field has the form
$$
ds^2= \left(1+ \dt\frac{(\kappa c^2\rho -2\Lambda ) r^2}{6}
\right)dt^2
 -\frac{dr^2}
{1 - \dt\frac{(\Lambda + \kappa c^2\rho )r^2}{3} }
- r^2(d\theta^2 + \sin^2\theta\cdot d\varphi^2 )
$$
In topos ${\bf Set}^{\bf L_{op}}$ at stage {\bf 1} this metric
complies with the metric of the Minkowski space-time. Roughly speaking,
non-classical "dust" vacuum  has the "infinitesimal" weak
gravitational field.

{\bf 2.2. Field of gas ball.} Suppose that gravitational
field was created by
spherical gas ball of radius $a$ with tensor of the energy
$\tilde{T_{ik}}$. From formula (5) under condition of
absence a singularity of  matter of the form $\lambda |_{ r=0} =0$ we
get
$$
\lambda=- \ln \left(1-\dt\frac {\kappa}{r}\int\limits_{0}^{r}
 \tilde{T_0^0} r^2 dr
 - \dt\frac{\Lambda r^2}{3} \right)
$$
Outside of the ball we have the vacuum with $\tilde{T_{ik}}=c^2\rho u_iu_k$
and with gravitational field that was studied in the preceding point.
So it is possible to use expression (13), from which it follows that
$$
\lambda= -\ln\left(1-\dt\frac{\Lambda
+ \kappa c^2\rho}{3}r^2 +\dt\frac{C}{r}\right)
$$
Comparing both expressions under $r=a$, we find that
$$
C=\kappa\cdot\left(\frac{c^2\rho a^3}{3} -
\int\limits_{0}^{a}\tilde{T_0^0} r^2 dr\right)\eqno(18)
$$

By using  that  $C$ and $\rho$  are inconvertible and (18), we get
that $\int_{0}^{a}\tilde{T_0^0}r^2 dr$ is inconvertible. This is possible
only in two cases: 1) $\tilde{T_0^0 }$  is inseparable from the zero;
2) $a$ is inseparable from the zero.

Thereby, the following theorem is true.

{\bf Theorem.} {\it Let gas ball possesses classical nonzero density
($\tilde{T_0^0 }\neq 0)$ and creates external spherically symmetrical
gravitational field (17) with dust infinitesimal density $\rho$.
 Then ball has infinitesimal sizes.
}

It is interesting that in the classical case the  Schwarzschild solution
was found in the suggestion that gravitational field is created by the
ball that is so naming material point, which is not having sizes.
Such situation was characterized by the word "simplification".
In non-classical case a material point gets
wholly legal sizes, but they will be described
infinitesimal numbers.

Notice that unlike classical solution, constant $C$ can not so simply
be expressed through the mass of ball. Really, following classical
procedure, we are noting that on greater distances, where field
is weak, the field must be  described by the Newton's Law.
Hence, $g_{00}= 1-\dt\frac{2Gm}{c^2r },$ where
$m$ is a mass of ball. On the other hand,
$g_{00}=1+\dt\frac{\kappa c^2\rho - 2\Lambda}{6} r^2 + \dt\frac{C}{r}.$
Thence it is seen that
$C=\dt\frac{2\Lambda - \kappa c^2\rho }{6} r^3 - \dt\frac{2Gm}{c^2}$. This
gives contradiction with $C=const.$

In topos ${\bf Set}^{\bf L_{op}}$ at stage {\bf 1} metric (17)
complies with the metric of the Partial theory of Relativity.
So cosmological  model with this metric can be called  a generalized
model of the Partial theory of Relativity .

\end{document}